\documentclass[pre,preprint]{revtex4}
\usepackage{amsmath,amssymb,graphicx}
\usepackage{latexsym}
\usepackage{color}
\usepackage{graphicx}

\begin{document}

\title{Large scale fluctuations and dynamics of the Bullard - von K\'arm\'an dynamo }

\author{ Gautier Verhille, Nicolas Plihon, Gr\'egory Fanjat,  Romain Volk, Mickael Bourgoin,  and Jean-Fran\c{c}ois Pinton}
\affiliation{Laboratoire de Physique de l'\'Ecole Normale Sup\'erieure de Lyon, \\  CNRS \& Universit\'e de Lyon, F-69364 Lyon, France}

\begin{abstract}
{A synthetic fluid dynamo built in the spirit  of the Bullard device~\cite{Bullard} is investigated. It is a two-step dynamo in which one process stems from the fluid turbulence, while the other part is an alpha effect achieved by a linear amplification of currents in external coils~\cite{BVK1}.  Modifications in the forcing are introduced in order to change the dynamics of the flow, and hence the dynamo behavior. Some features, such as on-off intermittency at onset of dynamo action, are very robust. Large scales fluctuations have a significant impact on the resulting dynamo, in particular in the observation of magnetic field reversals.} 
\end{abstract}

\maketitle

\section{Introduction}
$\alpha - \omega$ dynamo models are extensively used in the description of natural dynamos~\cite{astroDyn}. In these approaches, the $\omega$- effect converts a poloidal field into a toroidal component, while the $\alpha$-effect is invoked for the conversion back from toroidal to poloidal. In planets, the electrically conducting fluid is a molten metal (iron for the Earth), {\it i.e.} a fluid with a very low magnetic Prantl number. As a consequence, turbulence has to be fully developed for induction effects to be able to overcome (Joule) dissipation, and hence induce a dynamo. Turbulent dynamos have been extensively studied in the context of mean-field hydrodynamics~\cite{MoffattBook,RadlerMF}. In this approach, the turbulent small scale induction processes contribute to the dynamics of the large scale magnetic field through the mean electromotive force $\overline{\epsilon}_i = \alpha_{ij}  \overline{B_j} + \beta_{ijk} \partial_j \overline{B_k}$ (the reader is referred to~\cite{Stepanov2006} for generalised expressions). The $\alpha$ and $\beta$ tensors depend on the small scale turbulence, and expressions for them have been established in several contexts. Measurements, however, have revealed that their contribution is small, at least at low magnetic Prandtl numbers and moderate magnetic Reynolds numbers~\cite{alphaPerm,betaPerm}. An $\alpha$ contribution can also originate from large scale motions through Parker's stretch and twist mechanism~\cite{Bourgoin04,VKSalpha}.  Recently, the observation of a self-sustained axisymmetric magnetic field  in the VKS experiment~\cite{P1} has been modeled as an $\alpha$-$\omega$ dynamo~\cite{gafd,nore,Pcras,P5}. This experiment has also revealed a rich time-dynamics of the magnetic field~\cite{P2,P3,P4}, raising issues concerning the contribution of flow fluctuations on the dynamo behavior.
It has been shown indeed in several studies that von K\'arm\'an flows display large fluctuations at very low frequency in time ~\cite{VKBifurc1,VKBifurc2,torre2007} which have a leading role on magnetic induction characteristics~\cite{Volk06}.

In the present study we consider experimental synthetic dynamos~\cite{BVK1} in which an $\alpha$-effect is modeled by means of an external wiring, while a second induction process originates from the flow itself and takes into account the full turbulent fluctuations. One of these synthetic dynamos is an $\alpha-\omega$ dynamo similar to the one described in~\cite{BVK1}. The $\omega$ effect is linked to the flow differential rotation and has been extensively studied in previous experimental~\cite{Odier,Volk06} and numerical works~\cite{Bourgoin04} : if a magnetic field is applied parallel to the cylinder axis by a set of coils, the fluid differential rotation  induces a toroidal magnetic field. Since the flow is fully turbulent (with integral kinetic Reynolds exceeding $10^5$), the induced toroidal magnetic field is highly fluctuating: the $\omega$-effect incorporates the flow turbulence. An effective $\alpha$-effect is then achieved when the intensity of the induced toroidal field is used to control a linear amplifier driving the current in the coils. This feed-back loop generates a dynamo.

The main goal of the present study is to characterize how the resulting dynamo bifurcation and magnetic field dynamics are affected when large scale properties of the von K\'arm\'an flow are modified by the insertion of various appendices inside the vessel~\cite{RaveletThesis}, or when rotating the impellers at unequal rates -- as a way to impose global rotation onto the flow~\cite{VKBifurc1}.

The experimental set-up is described in details in section \ref{secsetup}. The dynamics of the synthetic Bullard-von K\'arm\'an dynamos are described and analyzed in section III. Conclusions are drawn in section IV.

\section{Experimental setup}\label{secsetup}
\subsection{von K\'arm\'an gallium flow}
The synthetic Bullard-von K\'arm\'an dynamo is built upon a von K\'arm\'an flow. This flow is produced by the rotation of two impellers inside a stainless steel cylindrical vessel filled with liquid gallium. The cylinder radius  $R$ is 97~mm and its length is 323~mm. The impellers have a diameter equal to 165~mm and are fitted to a set of 8 blades with height $10$~mm. They are separated by a distance $H =203$~mm. The impellers are driven by two AC-motors  which provide a constant rotation rate in the interval $(F_1, F_2) \in [0.5, 25]$~Hz. In most of the cases, the flow is driven by symmetric forcing at $F_1=F_2=F$. The system  is cooled by a water circulation located behind the driving impellers; the experiments are run at a temperature between $40^{\circ}$C and $48^{\circ}$C. Liquid gallium  has density $\rho = 6.09 \times 10^{3} \; {\rm kg}{\rm m}^{-3}$, electrical conductivity  $\sigma = 3.68 \times 10^{6} \; {\rm ohm}^{-1}{\rm m}^{-1}$, hence a magnetic diffusivity $\lambda = 1/\mu_0\sigma = 0.29 \; {\rm m}^2 {\rm s}^{-1}$. Its kinematic viscosity is  $\nu = 3.1 \times 10^{-7}  {\rm m}^{2}{\rm s}^{-1}$.  The integral kinematic and magnetic Reynolds numbers are defined as ${\rm Re} = {2\pi R^{2}F}/{\nu}$ and ${\rm R_m} = 2\pi R^{2}F/\lambda$. Values of ${\rm R_m}$ up to $5$ are achieved, with corresponding ${\rm Re}$ in excess of $10^6$.

Magnetic induction measurements are performed using Hall sensor probes inserted into the flow in the mid plane.  Data are recorded using a National Instrument PXI-4472 digitizer at a rate of $1000$~Hz with a 23 bits resolution.

\begin{figure}[ht!]
\begin{center}
\includegraphics[width=12cm]{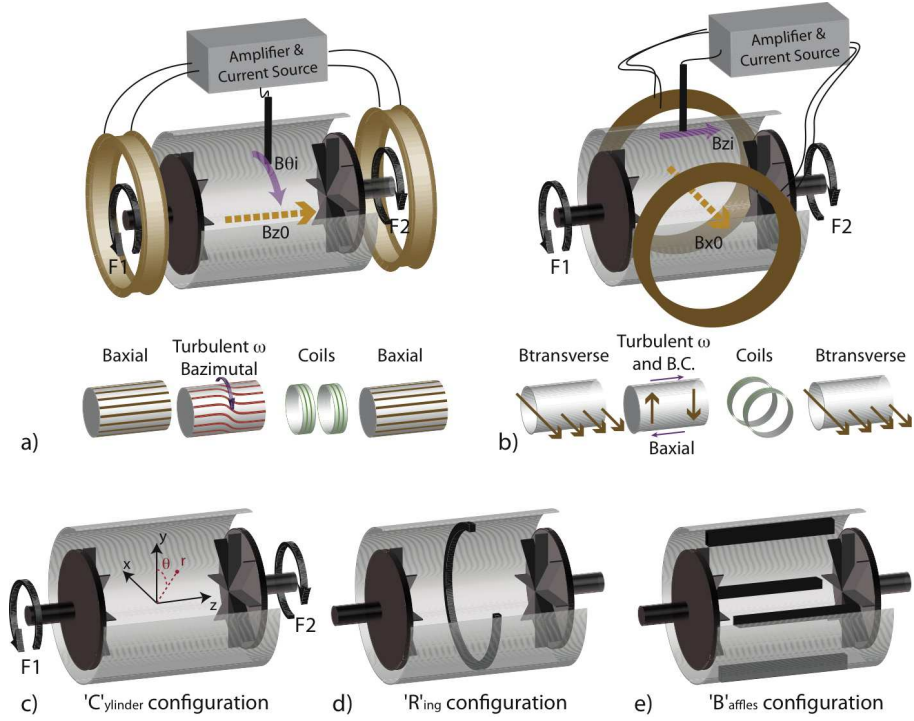}
\end{center}
\caption{Experimental setup : a von K\'arm\'an flow of liquid gallium is generated in a cylindrical vessel between two counter rotating impellers driven by two AC-motors. (a) Synthetic $\alpha-\omega$ dynamo (axial dipole), the turbulent induced azimutal field is measured at 0.56 R in the mid-plane between impellers. (b) Synthetic $\alpha-BC$ dynamo (equatorial dipole), the turbulent induced axial field is measured at 0.87 R in the mid-plane between impellers. (c) Cylinder ($C$) configuration of the vessel and coordinate axis. (d) Ring  ($R$) configuration and (e) Baffles ($B$) configuration. }
\label{VKgeom}
\end{figure}

\subsection{Dynamo loops}
{\it Axial dipole}\\
A first dynamo configuration which generates an axial dipole is shown in figure~\ref{VKgeom} (a). As the flow is forced by counter-rotating impellers, the differential rotation in the fluid motion induces an azimutal magnetic field $B_{\theta}^{I}$ when an applied axial magnetic field $B_{z}^{A}$ is imposed by axial coils; this is the $\omega$-effect~\cite{MoffattBook}. This induced azimuthal field is measured from a local probe in the mid plane at $r = 0.56 R$ and its value is used to drive the current source which feeds currents into the axial coils. This external amplification constitutes the $\alpha$ part of the dynamo cycle, since an axial field stems from an azimuthal one. The feed-back amplification is linear ($I_{\rm coils} \propto B_{\theta}^{I}$ up to a saturation value $I_{\rm coils}^{\rm sat}$, which is the maximum current that can be drawn from the current source).  The main component of the resulting dynamo is an axial dipole.

For all available values of the rotation rate of the driving impellers, the magnetic Reynolds number is low enough to be in a quasi-static approximation. In this regime,  $B_{\theta}^{I} \sim (\Delta^{-1}/\lambda)B_{z}^{A} \partial_z v_\theta$, where $v_\theta$ is the azimutal velocity~\cite{Bourgoin04}. The $\omega$ part of the dynamo cycle is thus very much dependent on the statistical properties of the velocity gradients at all scales. \\

{\it Equatorial dipole}\\
Another possibility is to arrange the external coils so that they generate a field perpendicular to the rotation axis, say $B_x$ (an equatorial dipole). In this case the induced magnetic field components which drive the current source is the axial induction $B_z^{\pi/2}$ measured in the mid plane at $90^\circ$ from the applied field. Numerical studies~\cite{Bourgoin04b} have shown that, in the quasi-static approximation, $B_z^{\pi/2}$ is mainly induced from $B_x$ and the radial vorticity $\omega_r$ of the flow via a boundary-condition effect due to the jump in electrical conductivity at the wall: $B_z^{\pi/2} \sim - (\nabla \times )^{-1} (\Delta^{-1}/\lambda)([\lambda] {B}_0.\nabla)\omega_r$, where  $ \left[ \lambda \right]$ is the jump in magnetic diffusivity at the vessel boundary.
In the following we will refer to this mechanism as $BC$-effect, for \textit{Boundary-Conditions effect}, and the equatorial dipole dynamo is of $\alpha-BC$ type.

In von K\'arm\'an flows, the $BC$-effect is particularly intense because large scale radial vortices develop in the mid plane as a result of a Kelvin-Helmholtz like instability in the shear layer~\cite{VKBifurc2,torre2007}. These radial vortices are moreover known to have a highly fluctuating dynamics which in turn generates strong fluctuations of the $BC$ induced field. \\

{\it Implementation}\\
For each configuration, the gain $G$, of the linear amplifier which imposes the current output in the coils from the induced magnetic field value is kept constant. The control parameter being the rotation frequency of the impellers $F_i$, we adjust the value of $G$ so that the critical rotation frequency for dynamo onset is $F_c=8.3$ Hz. This gain is based on time-averaged values of the magnetic induction when a fixed current is imposed in the external coils -- an `open-loop', or induction  configuration. One first measures the efficiency $A_\text{eff}$ of induction at the probe location by $\langle B^I \rangle_t = A_\text{eff} F \langle B^A \rangle_t$, where $\langle . \rangle_t$ is the time-averaged operator. Then one sets the gain $G$ of the closed-loop conversion (dynamo configuration), $B^A = G B^I$ such that the estimated dynamo threshold $F_c = 1/ G A_\text{eff}$ is constant and equal to 8.3~Hz. In other words, the gain is adjusted such that, for a given flow forcing parameter $F$, the time-averaged value of the amplified induced field $G\langle B^I \rangle_t$ is constant whatever the flow configuration and dynamo type.

Saturation is set by the maximum current available from the source; the saturated magnetic field value is of the order of a few tens of  Gauss. As a result, the Lorentz force is too weak to react back onto the flow field (the interaction parameter is less than $10^{-2}$). The synthetic Bullard-von K\'arm\'an is thus similar to a kinematic dynamo.

\subsection{Flow configurations}
The scope of the present study is to investigate the influence of the flow dynamics on the behavior of the self-sustained magnetic field. Modifications are achieved by changing the flow, either by inserting appendices to the inner wall of the cylinder or by driving the flow assymetrically ({\it i.e.} rotating the impellers at different speeds).

Three geometric configurations have been tested. They are shown in figure~\ref{VKgeom}(c,d,e). In the base configuration (hereafter called $C$ for Cylinder) there is no attachement. In configuration $B$, four longitudinal baffles are attached; they are rectangular pieces 150 mm long, 10 mm tall and 10 mm thick, mounted parallel to the cylinder axis. In configuration $R$, a ring is attached in the mid-plane; it is 15 mm tall and 4 mm thick. The respective influence of these attachments have been studied in water experiments~\cite{RaveletThesis} where it was found that they modify significantly the dynamics of the differential shear layer in the mid-plane. The presence of the ring or the baffles tends to stabilize the central shear layer; they modify also the number and time evolution of the radial vortices in the shear layer. 

Similarly, when the flow is driven by counter-rotating the impellers at different speeds, the flow bifurcates between 2-cells and 1-cell configurations~\cite{VKBifurc1,torre2007}, hence leading to substantial modifications of the flow turbulence and large scale topology. The unbalance in the driving of the flow is measured by the ratio $\Theta = (F_1 - F_2)/(F_1 + F_2)$, $\Theta =0$ corresponding to exact counter-rotation and $\Theta = 1$ to the fluid being set into motion by the rotation of one impeller only. The transition occurs at a critical value of $\Theta$ above which the shear layer abruptly moves from the mid-plane to the vicinity of the slower impeller (the flow transits then from a two counter-rotating cells geometry to a single cell geometry). It is associated with an abrupt change in the mechanical power required to keep the impellers rotating at fixed rates.

Altogether, by inserting appendices in the flow or driving it asymmetrically, one has the possibility to vary the mean profiles and fluctuation characteristics of the von K\'arm\'an flow. The impact of these changes on the synthetic $\alpha-\omega$ and $\alpha$-BC dynamos is studied in detail below. Note that though both $\omega$ and $BC$ effects are primarily generated by the same differential rotation, they are sensitive to different topological properties, with also different dynamical behavior of the mid-plane shear layer:  the $BC$-effect is sensitive to the radial vorticity ($\omega_r$) in the vicinity of the vessel wall while the $\omega$-effect is sensitive to the axial change of the toroidal velocity in the bulk of the flow. The natural fluctuations of these flow gradients are known to be of different kinds : radial vorticity fluctuations are closely linked to the dynamic of strong intermittent Kelvin-Helmhotz vortices in the shear layer, while the axial gradient of toroidal velocity fluctuates according to local and global displacements of the shear layer.  Exploring both the axial and equatorial dipoles configuration give us then a way to probe selectively the impact of these velocity gradients on the dynamo.

\section{Dynamo and flow dynamics }
\subsection{Axial dipole, $\alpha-\omega$ dynamo}\label{secaxial}
In this section, the flow is driven symmetrically: impellers counter rotate at the same rate $F = F_1 = F_2$. The external coils are set to generate an axial dipole field. The turbulent $\omega$-effect (the source of the fluctuations for the dynamo) is first analyzed as an induction process on its own. Dynamical features of the $\alpha-\omega$ dynamos are then described and analyzed.

\begin{figure}[ht!]
\centerline{\includegraphics[width=13cm]{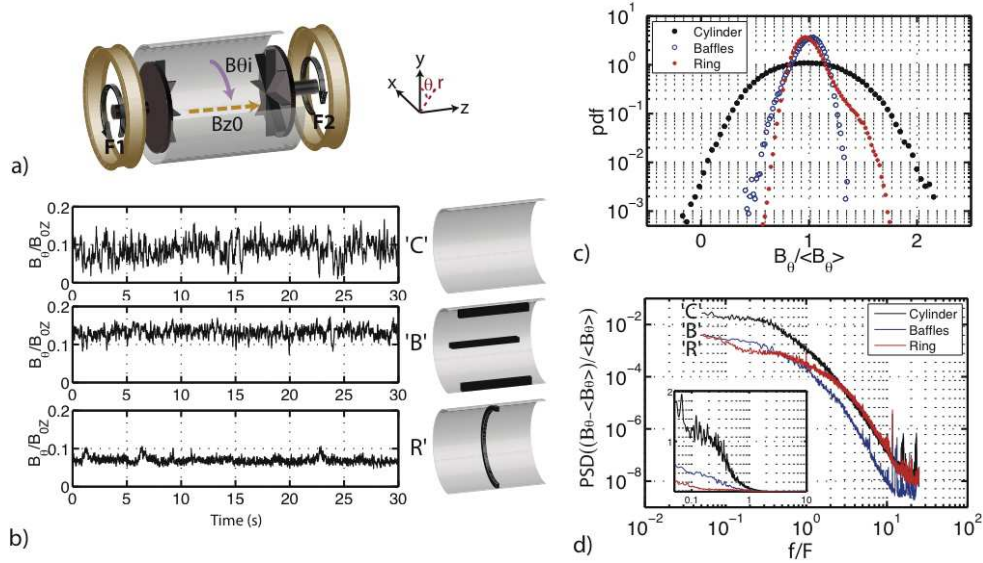}}
\caption{Axial applied field, azimutal induction features at r = 0.56 R. (a) Experimental setup, (b) Temporal dynamics for the three flow configurations. (c) Probability density functions and (d) Power spectrum density of the azimutal induced field (log-lin representation in the inset). Note that, for the last two graphs, the induced fields are normalized by their time-averaged values.}
\label{axialinduce}
\end{figure}

\subsubsection{Turbulent $\omega$ induction}
We first consider induction measurements obtained when a constant axial magnetic field is applied.  We show the temporal dynamics of the induced field in figure~\ref{axialinduce}(b), and display the probability density functions (PDFs) and time spectra of $B_\theta/\langle B_\theta \rangle$ in figure~\ref{axialinduce}(c-d).
Note that in order to compare the fluctuations for the different configurations, we have normalized the induced field by its mean amplitude in figure~\ref{axialinduce}(c,d). Indeed, in the dynamo loop, the value of the gain $G$ is adjusted so that the time-averaged amplified  induced field for a fixed rotation rate is the same whatever the flow configuration.

\begin{itemize}

\item[$\bullet$] $C$ configuration. The shear layer is fully developed in this less constrained state. As a result, its motion explores a large portion of the flow volume. These motions result in large variations of the amplitude of the induced field, with a quasi-Gaussian statistics. The power spectrum of fluctuations in time reveals an important low frequencies dynamics. These features have been discussed in detail in previous studies~\cite{Volk06}.

\item[$\bullet$] $R$ configuration. When a ring is installed in the mid-plane, the shear layer is pinned~\cite{RaveletThesis}. The axial gradient of the time-averaged azimutal velocity is peaked at the ring position \cite{RaveletThesis} and higher than in the $C$ configuration. The time-averaged azimutal induced field is nevertheless of the same order of magnitude in both configurations (cf. figure~\ref{axialinduce}(b). This is because the magnetic field `filters' the velocity gradients at a length scale of the order of the magnetic diffusion length $\eta_B \sim \sqrt{\lambda/F}~5$~cm at $R_m = 2$. The magnetic field is more sensitive to the difference of velocity on either side of the shear layer rather than in the actual slope of the velocity profile.

Another effect is that fluctuations in the azimuthal induction are reduced due to the shear pinning. The time spectra for the two configurations are similar in the inertial range, but the presence of the ring reduces low frequency fluctuations. The slight peak at very low frequency is attributed to a `depinning' occuring at irregular time intervals, also observed as a bimodality in the probability density function in figure~\ref{axialinduce}(c).

\item[$\bullet$] $B$ configuration. When baffles are inserted in the vessel, vortices of the shear layer are confined  to inter-baffles position and fluctuations of the shear layer are also damped~\cite{RaveletThesis}. Like in the $R$ configuration, the time-averaged induced field is weakly modified, but fluctuations are highly reduced as compared to the  $C$ configuration.\\
\end{itemize}

For these three configurations, the mean induced field and normalized fluctuations for $R_m = 2$ are the following: $\langle B_\theta^I/B_z^A\rangle = 0.08; 0.12; 0.07$ and $(B_\theta^I)_\text{rms}/ \langle B_\theta^I\rangle =  0.32; 0.11; 0.1$ for the $C$, $B$ and $R$ configurations respectively. Comparison of the spectra~\ref{axialinduce}(d) show that modifications of the flow strongly modify the low-frequency, large scale dynamics of the turbulent induction, while keeping small-scale (inertial range)  dynamics almost unchanged. 

\subsubsection{Dynamo: axial dipole}
\begin{figure}[ht!]
\centerline{\includegraphics[width=13cm]{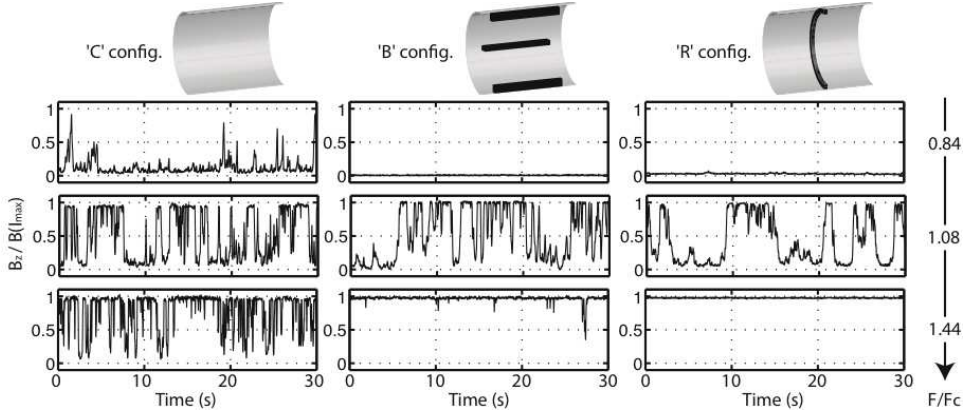}}
\caption{Axial dipole dynamo temporal features: temporal evolution of the dipole amplitude as a function of driving frequency for the three configurations.}
\label{axialdyn}
\end{figure}

We then analyse the behavior of the synthetic $\alpha-\omega$ dynamo ({\it i.e.} when the measured azimuthal field actually drives the current in the induction coils).  Time signals of the axial magnetic field (normalized to the maximum field)  are shown in figure~\ref{axialdyn} for the three configurations and an increasing forcing parameter F (from top to bottom). For the three configurations and whatever the value of the forcing parameter, the axial dipole dynamo is always observed homopolar: when growing in either polarity, the field remained of same polarity. However both polarities were observed for different realizations ({\it i.e.} when stopping the motors and increasing the rotation rate above the critical value), as is expected from the $\mathbf{B} \rightarrow -\mathbf{B}$ symmetry of the MHD equations.

Let us first analyze the $C$ configuration, for which last section showed that fluctuations of the $\omega$-effect are the highest. At very low rotation rates $F \ll F_c$, the magnetic field is null. When the rotation rate increases and approaches the threshold ($F\lesssim F_c$), the magnetic field remains null most of the time, but exhibits intermittent bursts. The density and duration of the dynamo state increase rapidly with the impellers rotation rate: as $F >1.3 F_c$, the  magnetic field in mostly non-zero, with intermittent extinctions for short time intervals. The system thus bifurcates from a non dynamo to a dynamo state via an on-off intermittency scenario, also previously observed in numerical models of the dynamo~\cite{SweetPP8,AlexakisPRE77}. This kind of bifurcation is characterized by specific features of the signal: its probability density function is peaked at zero near onset and then decays algebraically; the PDFs of `off' times scales as $T_{\rm off}^{-3/2}$~\cite{HeagyPRE49}. 

\begin{figure}[ht!]
\centerline{\includegraphics[width=13cm]{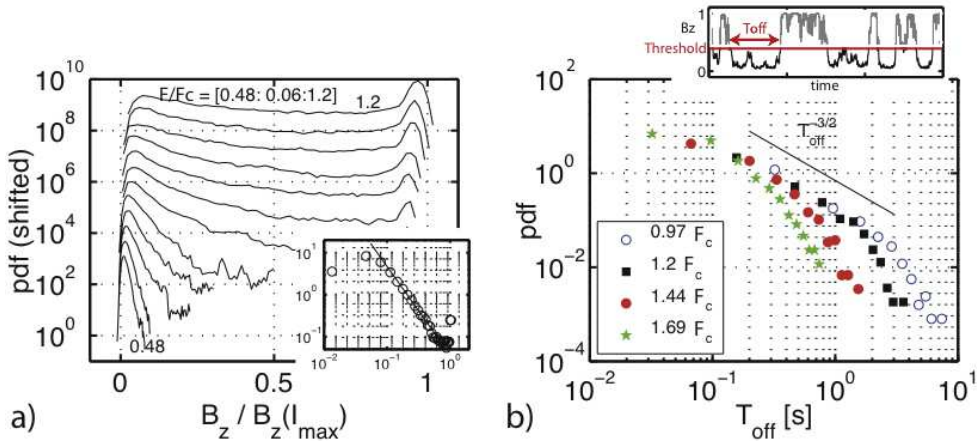}}
\caption{Axial dipole dynamo for the $C$ configuration: (a) evolution of the probability density function of the axial magnetic field as a function of the driving parameter - inset: log-log plot for $F = 0.84 F_c$- and (b) - inset: definition of inter-burst time interval - main: evolution of the probability density function of the inter-burst intervals as a function o the driving parameter.}
\label{axialdynonoff}
\end{figure}

Evolution of the measured PDFs of magnetic field as a function of the forcing parameter  is shown in figure \ref{axialdynonoff}(a). The PDF is peaked at zero for rotation frequencies up to $1.2 F_c$, and extends to large values, even at low rotation frequencies. A log-log representation is shown in inset of figure~\ref{axialdynonoff}(a) and shows a typical negative power-law for the PDF (with an exponent proportional to the control parameter). This feature was observed in numerical simulation of on-off models~\cite{AumaitreJSP123,AumaitrePRL95} and simulations of MHD equations~\cite{SweetPP8}. On-off intermittent signals also display an universal behavior of inter-burst time intervals, or time interval between dynamo activity, named $T_\text{off}$  - for 'off' phases or laminar phases. This time interval is defined in the inset of figure~\ref{axialdynonoff}(b): one sets a threshold and computes the time $T_\text{off}$ for which the dynamo field is below this threshold.  The probability density functions of  $T_\text{off}$ are displayed in figure~\ref{axialdynonoff}(b) for four values of the forcing parameter. These probability density functions have a power law dependence with a cut-off at large values (the distribution of the `off' phases has been checked to be independent of the value of the threshold). A $T_\text{off}^{-3/2}$ scaling is shown for comparison with the experimental data, showing that the data statistics is in agreement with the -3/2 power law scaling behavior predicted for on-off intermittency\cite{HeagyPRE49}, observed in numerical simulations~\cite{AumaitreJSP123,AumaitrePRL95,SweetPP8} or in experimental investigations in electronic circuits~\cite{HammerPRL73}, gas-discharge plasmas~\cite{FengPRE58} or convection in liquid crystals~\cite{JohnPRL83}. Corrections to the power-law scaling are not obvious, so that more complex scenarii (such as ` in-out' intermittency~\cite{CovasChaos11}) may not apply here.

The variation of the cut-off time in the PDF of $T_{\rm off}$  with the control parameter is steep. We observe a power law with a $-5$ exponent in a small range  $F\in [0.97F_c,1.69F_c]$. This is in contrast with numerical  models with multiplicative noise for which a decay as $F^{-2}$ is quoted~\cite{AumaitreJSP123}.

\begin{figure}[ht!]
\centerline{\includegraphics[width=12cm]{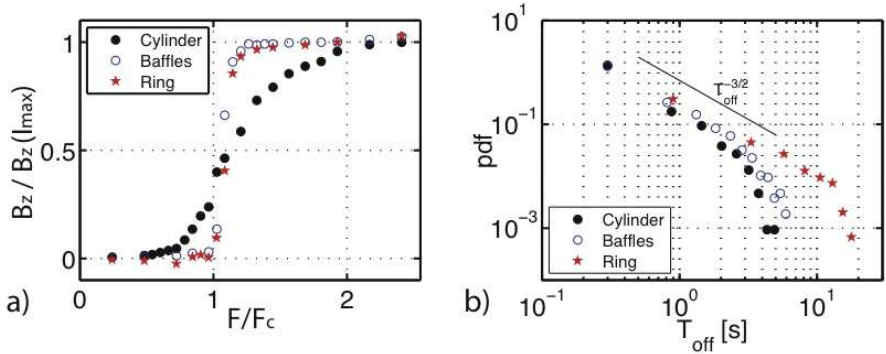}}
\caption{Features of the axial dipole dynamo: (a) evolution of the time-averaged axial magnetic field as a function of the rotation frequency for the three configurations; (b) probability density function of the inter-burst time interval for the three configurations at $F = 1.08 F_c$.}
\label{axialdynbif}
\end{figure}

Modifications of the $\omega$-effect with the other two configurations drastically change the dynamo features. As can be seen in figure~\ref{axialdyn}, for the $B$ and $R$ configurations, no dynamo bursts are observed for impellers rotation rates well below threshold and 'off' phases are no longer present well above threshold. The associated bifurcation curves are displayed in figure~\ref{axialdynbif}(a). The bifurcation is steeper for the $B$ and $R$ configurations as compared to the $C$ configuration, {\it i.e.} the width of the intermittent domain decreases when inserting appendices in the vessel. The on-off intermittent regime is nevertheless always observed in a very narrow range near onset. Figure~\ref{axialdynbif}(b) shows the probability density functions of the `off' phases for the three configurations near the critical rotation rate ($F = 1.08 F_c$). The -3/2 power law scaling is independent of the configuration, while the exponential cut off increases when the low frequency component of fluctuations decreases. The mean length of `on' phases also significantly increases when low frequency of the noise decreases, as can be observed in figure~\ref{axialdyn}.

As a partial conclusion, we observe that several features (such as the on-off scenario and waiting times statistics) of the axial dynamo at onset are robust with respect to large scale modifications of the von K\'arm\'an flow. Other features (such as the width of the transition and the frequency of incursions to or from dynamo states) are on the contrary strongly dependent on the large scale dynamics of the flow.

\subsection{Equatorial dipole, $\alpha-BC$ dynamo}\label{sectransverse}
We now consider the $\alpha$-BC dynamo configuration, generating an equatorial dipole field, perpendicular to the axis of rotation of the impellers. As for the axial case, we study the dynamo bifurcation in the $C$, $B$ and $R$ configurations, the flow being driven symmetrically with the impellers counter rotating at
the same rate $F = F_1= F_2$. As for the previous section, the gain of the feedback loop is adjusted for each configuration in order to keep constant the dynamo threshold estimated from time-averaged induction measurements ($F_c = 8.3$~Hz).

\subsubsection{Turbulent $BC$ induction}
Induction measurements are performed with a constant field, applied perpendicularly to the cylinder axis. The induced field in the axial direction is measured in the mid plane at $90^\circ$ from the applied field, at distance $0.87R$ from the axis. 

\begin{figure}[ht!]
\centerline{\includegraphics[width=12cm]{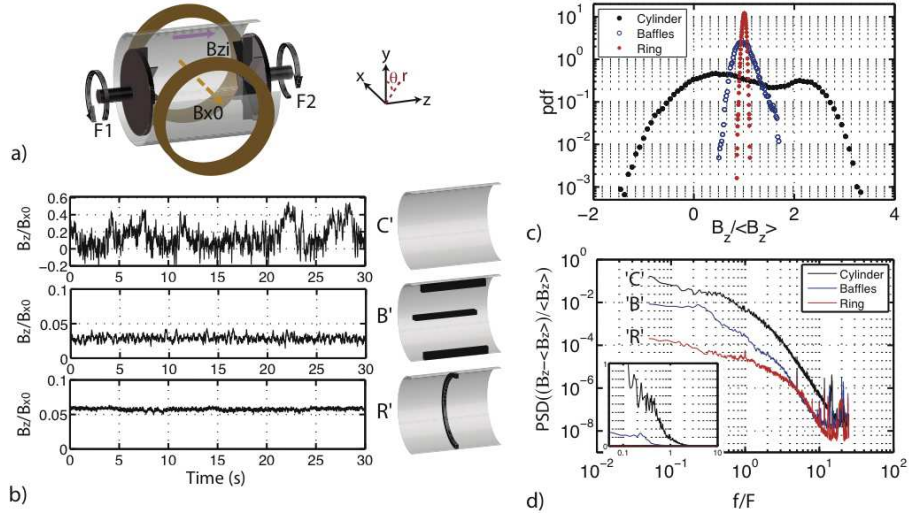}}
\caption{Transverse induction. (a) experimental set-up, (b) temporal evolution of axial induction $B_z^I(t)$ for a constant transverse applied field $B_x^A $, (note the difference in y-axis scales)  (c) probability density function and (d) time spectra of normalized induced fields - log-lin representation in the inset.}
\label{transind}
\end{figure}

\begin{itemize}
\item[$\bullet$] $C$ configuration. The induced field (figure~\ref{transind}(b)) shows a highly fluctuating dynamics. As seen in figure~\ref{transind}(c), the induced field distribution is bimodal, with a peak near zero induction and one at about twice the time-averaged value. The induction follows the dynamics of the radial vortices formed in the shear layer. A significant induced field is measured only when radial vortices sweep the probe. In between these events, the induced field is weak with large fluctuations -- the sign of the induced field is not prescribed. 

\item[$\bullet$] $B$ configuration. As compared to the above situation, the induced field is always non zero, a feature consistent with the observation~\cite{RaveletThesis} that the radial vortices are now confined between the baffles. A concurrent feature is that the fluctuations of induction are much reduced (a factor of 5 lower than the $C$ configuration). The spectra displayed in figure~\ref{transind} show that this reduction occurs at all scales.

\item[$\bullet$] $R$ configuration. In this case, the shear layer is pinned. The induction signal is again stationary. This is consistent again with observations~\cite{RaveletThesis,MonchauxThese} that the vortices are attached to the central ring. The fluctuations are reduced again by a factor of 5, as also seen in figure~\ref{transind}(c), this reduction lies essentially in the low frequency motions (figure~\ref{transind}(d)). 

\end{itemize}

For the three configurations, mean induced field and normalized fluctuations at $R_m=2$ are the following: $\langle B_z^I/B_x^A\rangle = 0.029; 0.029; 0.058$ and $(B_z^I)_\text{rms}/ \langle B_z^I\rangle =  0.90; 0.17; 0.03$  for the $C$, $B$ and $R$ configurations respectively.

\subsubsection{Dynamo: equatorial dipole}
\begin{figure}[ht!]
\centerline{\includegraphics[width=12cm]{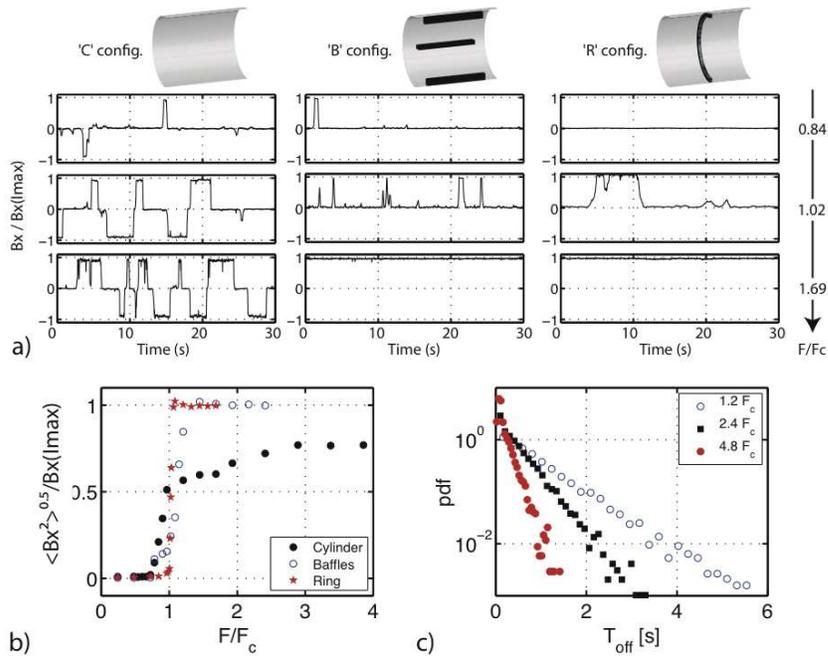}}
\caption{Equatorial dipole dynamo behavior: (a)time signals for increasing values of the rotation rate  of the impellers, from below to above threshold, (b) bifurcation curves and (c) probability density function of the 'off' phases for the $C$ configuration, {\it i.e.} the reversing dynamo.}
\label{transdynfeat}
\end{figure}

Time recordings of the equatorial dipole dynamo field are shown in figure~\ref{transdynfeat}(a). In each configuration, time signals are shown for three values of the rotation rates of the impeller, from below the critical frequency to above.

When compared to the previous axial dynamo cases, several features emerge. First in the absence of appendices in the vessel (configuration $C$), a dynamo field grows with either polarity and reverses spontaneously. Below threshold, bursts with both polarities $\pm {\bf B}_\text{eq}$ are observed. For higher values of the rotation rate, the fraction of time spent in a dynamo state increases. This is in sharp contrast with the $B$ and $R$ configurations for which homopolar dynamos are always generated after a rather abrupt bifurcation. In these configurations, bursts of dynamo action are observed near threshold but a steady state with a definite polarity is rapidly reached above threshold.

Let us describe and analyze in details the $C$ configuration for which the magnetic field exhibits reversals. The bifurcation curve, displayed in figure~\ref{transdynfeat}(b)-black shows a smooth transition from the null field regime to the dynamo regime. This transition occurs through bursts of both polarities below and around threshold. The time spent in the dynamo regime increases with the forcing parameter. As compared to all other Bullard-von K\'arm\'an dynamos obtained so far, a striking feature is that the reversing dynamo displays significant `off' phases even well above threshold. The probability density functions of the `off' phases for three values of the forcing parameter are shown in figure~\ref{transdynfeat}(c). The observed statistic is consistent with a Poisson distribution having a characteristic time scaling as $(F/F_c)^{-1}$. On the other hand, the distributions of the duration of the saturated ('on'; plus or minus) states (not shown) are consistent with a sum of two Poisson distributions with characteristic times in a ratio of 7 and scaling as $F/F_c$. Note that in this configuration, statistics are converged only when working with very large time recordings. Regimes at large $F/F_c$ have thus been obtained by increasing the gain $G$ in order to keep $F$ in the optimal range for operation with the liquid gallium kept at constant temperature (around $40^\circ$ C).

For the $B$ and $R$ configurations, the dynamics of the homopolar dynamo is very similar to the previous axial case. The on-off intermittent behavior is controlled by the low frequency content of the induction process: it is here restricted to a very narrow range of $F$ around $F_c$.

As a partial conclusion here, the equatorial dynamo also develops via an on-off intermittent regime, but some features have been significantly changed, such as the ability of the dynamo to undergo reversals of polarity. These reversals are only observed when strong fluctuations are present. This is a common feature of reversals dynamics when caused by stochastic transitions between the symmetric ${\mathbf B}$ and $-{\mathbf B}$ states~\cite{BenziPinton}.

\subsection{Global rotation}\label{secrotation}
In this section, we adress the case where the flow is driven with impellers counter-rotating at different rotations rates in the presence of a ring. This is an asymmetric $R$ configuration, as for the dynamical regimes of the VKS dynamo reported in~\cite{P2,P3,P5}. In order to emphasize the influence of the dynamics of the mid-plane shear layer, and its links with reversals, we focus on the equatorial dynamo loop.

\subsubsection{$BC$ induction with asymmetric forcing}
\begin{figure}[ht!]
\centerline{\includegraphics[width=14cm]{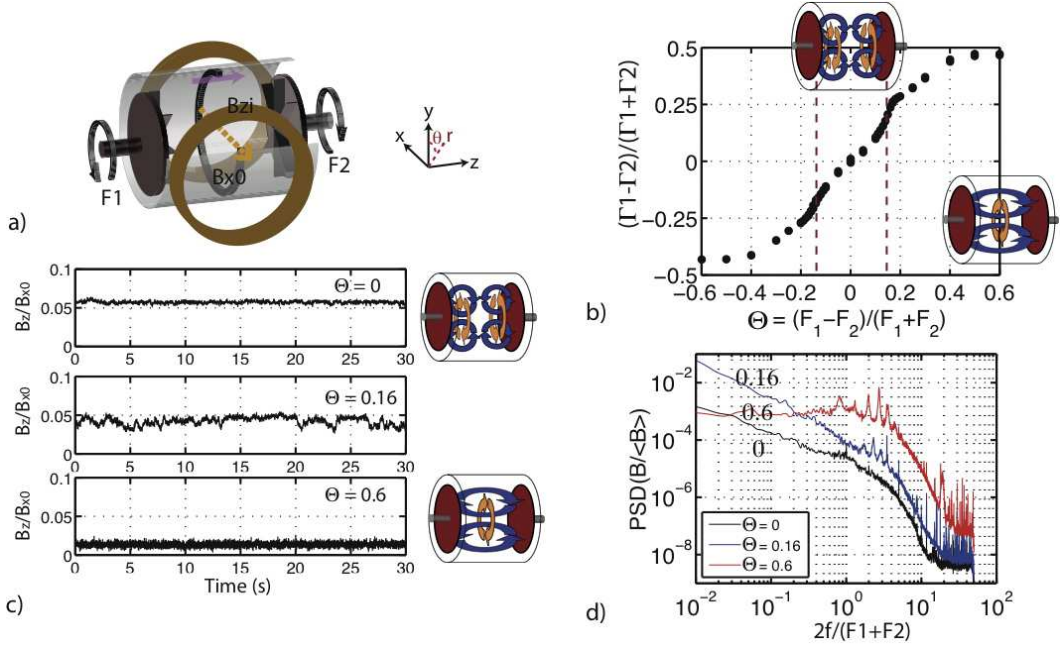}}
\caption{Influence of global rotation on $BC$ induction features: (a) time signals for three values of the global rotation, (b) dimensionless torque difference as the function of the flow asymmetry - insets show sketchs of the mean von K\'arm\'an flows, (c) time evolution of the induced field, and (d), corresponding power spectra.}
\label{globRot}
\end{figure}

As $\Theta = (F_1 - F_2)/(F_1 + F_2)$ is varied, one observes a clear transition in the torques $\Gamma_i$ driving the two impellers. Figure~\ref{globRot}(b) shows the dimensionless differences between the torques of the two impellers (these torques are deduced from the currents measured by the electrical drives that feed the motors). The bifurcation between a 2-cell flow and a 1-cell flow occurs at a criticl value $\Theta^c \sim 0.16$, in agreement with measurements made in water flows~\cite{VKBifurc1,NNN}. For $| \Theta | < \Theta^c$ the time-averaged flow consists of two main cells on either side of the mid plane, in which the toroidal and poloidal flows have opposite directions. Note that the regimes previously described correspond to a symmetric forcing, {\it i.e.} $ \Theta =0$. For $| \Theta | > \Theta^c$ the flow volume is dominated by one cell driven by the fast impeller, as is schematically drawn in figure~\ref{globRot}(b). It was shown in previous studies in water flows~\cite{VKBifurc1,RaveletThesis,NNN} that fluctuations of the flow diverge at the transition between the two flow regimes. 

Figure~\ref{globRot}(a) shows that fluctuations of magnetic induction strongly depend on the asymmetry of the forcing. Energy distribution across scales of the turbulent induced field also strongly depend on the asymmetry of the forcing as can be seen in  figure~\ref{globRot}(c).  Near $\Theta \sim 0.16$ (shear layer instability) the low frequency part of the spectrum is enhanced. At higher $\Theta$ values the spectrum is enhanced around and above the forcing frequency. One thus estimates that the flow has slow large scales fluctuations only near $\Theta^c$, while it has larger small scale fluctuations for $\Theta \gg \Theta^c$.

Evolutions mean induced field and normalized fluctuations at $R_m=2$ as a function of global rotation are the following: $\langle B_z^I/B_x^A\rangle = 0.058; 0.040; 0.013$ and $(B_z^I)_\text{rms}/ \langle B_z^I\rangle =  0.03; 0.16; 0.22$  for the $\Theta = 0; 0.16; 0.6$, respectively. 

\subsubsection{Asymmetric equatorial dynamo}
\begin{figure}[ht!]
\centerline{\includegraphics[width=12cm]{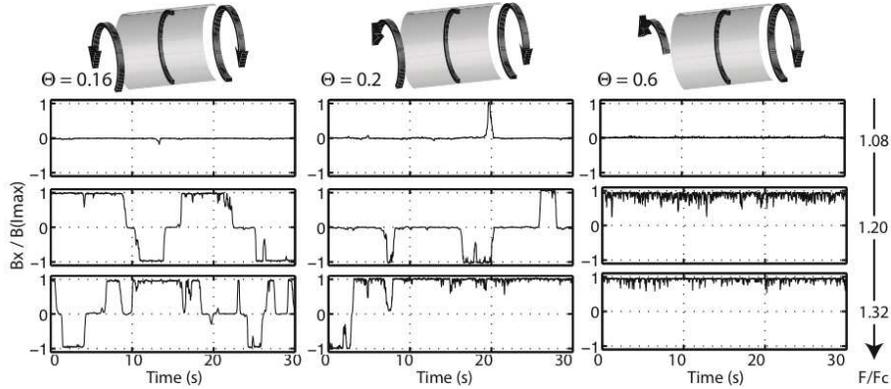}}
\caption{Transverse dynamo with global rotation: time recordings of the transverse magnetic field, for increasing values of $\Theta$ (left to right) and for increasing value of the mean  rotation rate $F$.}
\label{globDyn}
\end{figure}

 Time evolution of the equatorial dynamo field $B_x$ is shown in figure~\ref{globDyn} for three values of  $\Theta$ (refer to figure~\ref{transdynfeat} for the $\Theta=0$ case) and increasing values of the rotation rate $F$.

The $\Theta=0$ case, described in last section, showed an homopolar dynamo with an on-off intermittent regime at onset and a steep transition from the null-field regime to the dynamo regime. When increasing $\Theta$, a first observation is that both polarities occur, with chaotic reversals for $\Theta \sim \Theta^c$. This feature disappear for higher values of $\Theta$, {\it i.e.} when the flow has bifurcated toward the one-cell configuration, and one recovers an homopolar dynamo. The analysis of the $\Theta \sim \Theta^c$ regimes is of great interest since, for these regimes, the VKS dynamo displayed most of its dynamical regimes (chaotic reversals, symmetric and asymmetric bursts, oscillations...)~\cite{P2,P3,P5}. At the critical value $\Theta \sim \Theta^c$, the equatorial dynamo is bipolar, bifurcates through an on-off scenario and reverses chaotically. Close to the threshold value (the $\Theta = 0.2$ regime is displayed in figure~\ref{globDyn}), the dynamo also bifurcates via an on-off scenario and exhibits bursts of both polarities for a rotation rate close to $F_c$. At higher rotation rates, reversals occurences are less frequent and the dynamo eventually reaches an homopolar regime for $F/F_c>1.5$.

When the fow has bifurcated to the one-cell configuration, the equatorial dynamo is homopolar. On/off intermitency is observed in a very narrow range of driving frequencies.

\section{Concluding remarks}\label{secconcl}

In the present study, influence of the flow fluctuations on synthetic Bullard-von K\'arm\'an dynamos has been investigated. Two types of dynamo loops were studied: an $\alpha-\omega$ and a $\alpha-BC$ dynamo, where the $\omega$ and $BC$-effects incorporate turbulent fluctuations. Flow fluctuations modifications have been achieved by inserting appendices  in the vessel, or by driving the flow asymmetrically (for $\alpha-BC$ dynamo). Several robust features have been observed:
\begin{itemize}
\item The bifurcation occurs via an on-off intermittent regime at onset of dynamo action.
\item The on-off intermittent regime is controlled by the low frequency part of the fluctuating induction process considered in the dynamo loop: the higher the low frequencies of the fluctuations, the wider is the occurence of on-off intermittency among the control parameter.
\item For all studied configurations, the system spends half ot its time in the dynamo regime when driven at the critical forcing parameter defined on time-averaged induction processes.
\end{itemize}
However, some features strongly depend on the exact configuration. Reversing dynamos have been observed only for strong fluctuations of  the turbulent induction process. The low frequency part of the spectrum seems to play also a dominant role on the ability of the dynamo to reverse.

Hence, for this synthetic dynamo, salient features such as bursts of magnetic field activity or reversals are controlled by the underlying hydrodynamics. A detailed comparison with the dynamics of stochastic differential equations in the presence of noise is underway and will be reported elsewhere.

\vspace{15mm}
{\bf Acknowledgements. } This work has benefited from discussions with S. Auma\^itre,  E. Bertin, B. Castaing and F. P\'etr\'elis. It is supported by contract ANR-08-BLAN-0039-02.



\begin{thebibliography}{99}
\bibitem{NNN}
The impellers used to drive the water experiment in~\cite{VKBifurc1} were of a different geometry than in this present study. The disks were fitted with curved blades, while straight blades are used here.
\end{thebibliography}
\end{document}